\documentclass[10pt,conference]{IEEEtran}

\usepackage[nounderscore]{syntax}
\usepackage{booktabs}
\usepackage{multirow,makecell,multicol}
\usepackage{pifont}
\newcommand{\code}[1]{{\ttfamily \small #1}}

\usepackage{soul}

\usepackage{amssymb}
\usepackage{amsthm}
\theoremstyle{definition}

\usepackage[ruled,linesnumbered,noend]{algorithm2e}
\SetKwComment{Comment}{/*}{*/}
\RestyleAlgo{ruled}

\usepackage{caption}
\usepackage{subcaption}

\usepackage{listings}
\usepackage{xcolor}

\definecolor{codegreen}{rgb}{0,0.6,0}
\definecolor{codegray}{rgb}{0.5,0.5,0.5}
\definecolor{codepurple}{rgb}{0.58,0,0.82}
\definecolor{backcolour}{rgb}{0.96,0.96,0.96}

\lstdefinestyle{mystyle}{
    backgroundcolor=\color{backcolour},
    commentstyle=\color{codegreen},
    keywordstyle=\color{magenta},
    numberstyle=\tiny\color{codegray},
    stringstyle=\color{codepurple},
    basicstyle=\ttfamily\footnotesize,
    breakatwhitespace=false,
    breaklines=true,
    captionpos=b,
    keepspaces=true,
    numbers=left,
    numbersep=5pt,
    xleftmargin=8pt,
    showspaces=false,
    showstringspaces=false,
    showtabs=false,
    tabsize=2
}
\usepackage{xcolor}
\definecolor{celadon}{rgb}{0.67, 0.88, 0.69}

\usepackage{hyperref}

\usepackage{enumitem}
\setlist[itemize]{leftmargin=*}

\usepackage{tcolorbox}
\newenvironment{answerbox}{
\begin{tcolorbox}[colback=blue!5!white,colframe=blue!5!white,arc=0mm,grow to left by=1.5mm,left=0mm,grow to right by=1.5mm,right=0mm,top=0mm,bottom=0mm]
}
{
\end{tcolorbox}
}

\newcommand{\coverageQDiff}{6.1}
\newcommand{\coverageMorphQ}{8.1}

\newcommand{\methodName}{MorphQ}

\newcommand{\nMetamorphicRelationships}{ten}
\newcommand{\NMetamorphicRelationships}{Ten}
\newcommand{\nBugsFound}{nine}

\newcommand{\nDaysRun}{30}

\newcommand{\nBugReports}{13}
\newcommand{\nBugReportsMultipleTransformations}{8}

\newcommand{\percCrashesBenchmark}{23.2\%}
\newcommand{\totalTestedProgramPairs}{8,360}
\newcommand{\totalTestedProgramPairsAbbr}{8k}
\newcommand{\totalPairsWithDistrDifference}{56}

\begin{document}

\title{\methodName{}: Metamorphic Testing of the Qiskit Quantum Computing Platform}

\author{\IEEEauthorblockN{Matteo Paltenghi}
\IEEEauthorblockA{\textit{Department of Computer Science} \\
\textit{University of Stuttgart}\\
Stuttgart, Germany \\
mattepalte@live.it}
\and
\IEEEauthorblockN{Michael Pradel}
\IEEEauthorblockA{\textit{Department of Computer Science} \\
\textit{University of Stuttgart}\\
Stuttgart, Germany \\
michael@binaervarianz.de}
}

\maketitle

\thispagestyle{plain}
\pagestyle{plain}

\begin{abstract}
As quantum computing is becoming increasingly popular, the underlying quantum computing platforms are growing both in ability and complexity.
Unfortunately, testing these platforms is challenging due to the relatively small number of existing quantum programs and because of the oracle problem, i.e., a lack of specifications of the expected behavior of programs.
This paper presents \methodName{}, the first metamorphic testing approach for quantum computing platforms.
Our two key contributions are
(i) a program generator that creates a large and diverse set of valid (i.e., non-crashing) quantum programs, and
(ii) a set of program transformations that exploit quantum-specific metamorphic relationships to alleviate the oracle problem.
Evaluating the approach by testing the popular Qiskit platform shows that the approach creates over \totalTestedProgramPairsAbbr{} program pairs within two days, many of which expose crashes.
Inspecting the crashes, we find \nBugReports{} bugs, \nBugsFound{} of which have already been confirmed.
\methodName{} widens the slim portfolio of testing techniques of quantum computing platforms, helping to create a reliable software stack for this increasingly important field.

\end{abstract}

\makeatletter
\lst@AddToHook{PreSet}{\normallineskiplimit=0pt}
\makeatother

\section{Introduction}

Quantum software engineering is seeing an increasing interest from both academia and industry.
\emph{Quantum computing platforms}, such as Qiskit by IBM, Circ by Google, and Q\# by Microsoft, are for this emerging field what traditional compilers and execution environments are for traditional programs.
Ensuring the correctness of these platforms is crucial, since bugs in the platforms may undermine advances in algorithms and hardware.
A recent study~\cite{paltenghiBugsQuantumComputing2022} shows that quantum computing platforms are still plagued with bugs, many of which are due to quantum-specific bug patterns not present in traditional software.
The increasing importance of these platforms hence calls for automated testing techniques targeted at them.

Effectively testing quantum computing platforms currently faces two important challenges.
(C1) First, there currently are relatively few quantum programs, as the field is emerging and developers are only beginning to exploit its potential.
From a testing perspective, this means that test inputs are a scarce resource.
(C2) Second, another challenge is the well-known oracle problem~\cite{Barr2015}, i.e., not having a specification of the expected behavior triggered by an input.
Determining the expected behavior of a quantum program is particularly challenging since programs are composed of low-level operations, represented by gates, that translate to sometimes counterintuitive and highly abstract operations.

This paper presents \methodName{}, the first metamorphic testing approach targeted at quantum computing platforms.
The approach addresses challenge C1 by proposing the first automatic generator of quantum programs.
The generator combines template-based and grammar-based code generation to produce programs that use a diverse set of quantum gates and options for compiling and executing them.
To be effective, the generator carefully considers quantum-specific constraints, such as not applying any operation after a measurement gate because it would destroy the quantum state.
By respecting these constraints, the generator creates programs that are valid in the sense that they execute without crashing.

\methodName{} addresses challenge C2 through a novel set of \nMetamorphicRelationships{} metamorphic transformations.
Following the idea of metamorphic testing~\cite{Chen1998,chenMetamorphicTestingReview2018}, these transformations change a given source program into a follow-up program in such a way that the two programs have an expected output relationship, e.g., to be semantically equivalent.
If the expected output relationship does not hold, e.g., because the follow-up program crashes or otherwise changes the behavior, the approach reports a warning.
The metamorphic transformations are quantum-specific.
For example, they change the order of qubits, add null-effect operations by exploiting the reversible nature of quantum computation, partition a circuit that contains unrelated subcircuits, or change the set of hardware gates a program is compiled to.

Our evaluation applies \methodName{} to the popular Qiskit~\cite{QiskitQiskit2021} quantum computing platform.
During a two-day testing period, the approach generates, executes, and compares over \totalTestedProgramPairsAbbr{} pairs of quantum programs, many of which expose crashing bugs in the platform under test.
Manually inspecting a subset of the warnings reported by \methodName{}, we find and report \nBugReports{} bugs, \nBugsFound{} of which have been confirmed by the Qiskit developers so far.
For example, these bugs are caused by incorrectly implemented optimization passes, missing support for specific kinds of programs, and mistakes in exporting a program to QASM, an assembly-like language for quantum programs.

While testing traditional compilers has received significant attention~\cite{chenSurveyCompilerTesting2020}, we are aware of only one prior work, called QDiff~\cite{wangQDiffDifferentialTesting2021}, on automatically testing quantum computing platforms.
\methodName{} conceptually differs in multiple ways.
While QDiff starts from a small set of manually written programs, \methodName{} generates a large and diverse set of quantum programs from scratch.
Another difference is that QDiff is based on differential testing that compares executions with different optimization levels and backends, whereas we are the first to present metamorphic transformations for quantum programs.
Beyond conceptual contributions, we also empirically show our approach to complement prior work by finding previously undetected bugs and by reaching higher code coverage.

In summary, this work makes the following contributions:

\begin{itemize}
  \item A template-based and grammar-based program generator that creates valid quantum programs to use for testing purposes.
  \item \NMetamorphicRelationships{} quantum-specific metamorphic relationships to enable the first metamorphic testing framework for quantum computing platforms.
  \item Integrating the approach with the popular Qiskit platform and providing empirical evidence that \methodName{} reveals \nBugReports{} real-world, crashing bugs.
\end{itemize}

\section{Background on Quantum Computing}

Unlike classical computing, which is based on classical physics, quantum computing exploits the laws of quantum mechanics to perform computation.
Whereas in classical computing the minimal unit of information is a bit, which is either 0 or 1, in quantum computing the base unit is a \emph{qubit}, which can be a superposition of 0 and 1, representing a quantum state as $|\phi\rangle = \alpha|0\rangle + \beta|1\rangle$.
This superposition is manipulated along the computation and eventually, each qubit is \textit{measured} into either 0 or 1, with probabilities $|\alpha|^2$ and $|\beta|^2$, respectively.
Another important property is \textit{entanglement}, which means that the results of measuring two or more qubits are correlated.

\begin{figure}[t]
\begin{lstlisting}[language=Python, escapechar=\%]
# Create circuit
circ = QuantumCircuit(2)
circ.h(0)  # Hadamard gate  %\label{line:hadamard_gate}%
circ.cx(0, 1)  # Controlled not gate %\label{line:cnot_gate}%
circ.measure_all()
# Transpile for simulator
simulator = Aer.get_backend('aer_simulator')
circ = transpile(circ, simulator)
# Run and get counts
result = simulator.run(circ, shots=1024).result()%\label{line:simulator_1024_runs}%
counts = result.get_counts(circ)
# output: {'00': 530, '11': 494}
\end{lstlisting}
\caption{Example of a circuit to create entanglement.}
\label{fig:ex_quantum_program_code}
\end{figure}

\begin{figure}[t]
  \centering
  \includegraphics[width=0.47\textwidth]{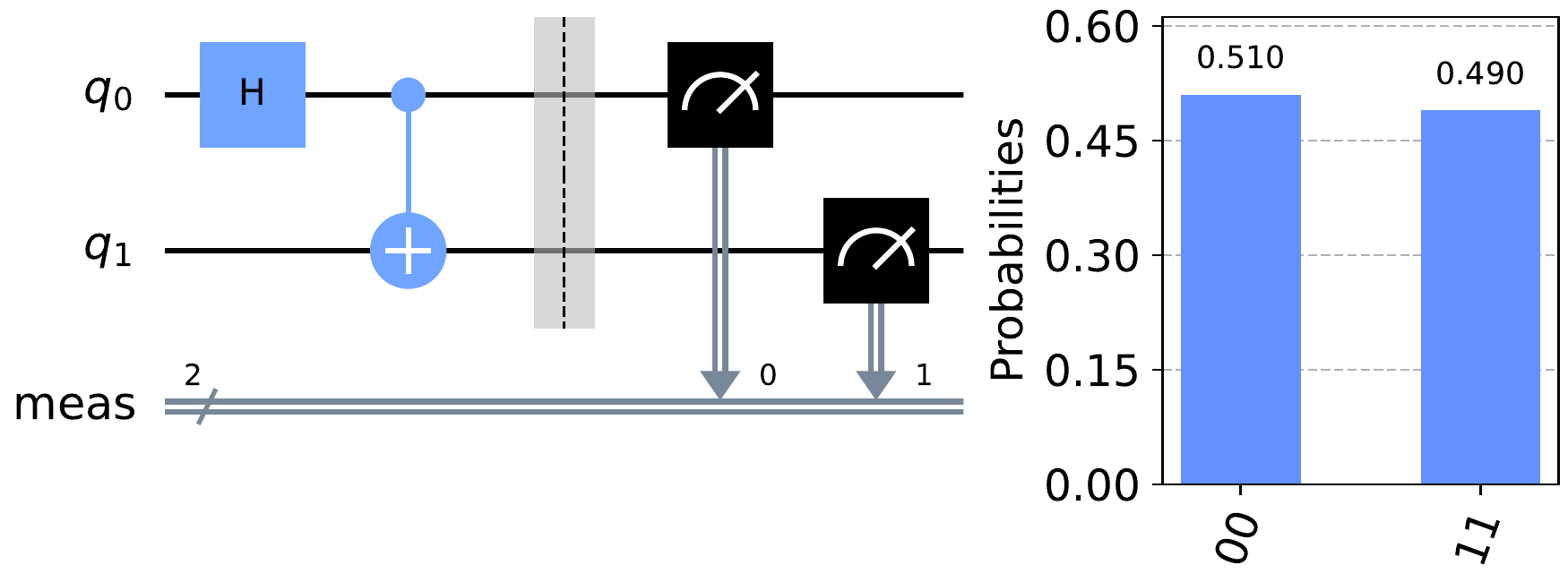}
  \caption{Visual representation of Figure~\ref{fig:ex_quantum_program_code} (left) and example of measurement result (right).}
  \label{fig:ex_quantum_program_draw}
\end{figure}

Figure~\ref{fig:ex_quantum_program_code} shows a simple quantum program, which creates an entanglement between two qubits.
The program applies a \textit{Hadamard gate} to the first qubit (line \ref{line:hadamard_gate}), which creates a superposition $|\phi\rangle = \frac{1}{\sqrt{2}}|0\rangle + \frac{1}{\sqrt{2}}|1\rangle$, and then a \textit{controlled not gate} (line \ref{line:cnot_gate}), which creates the entanglement between the first and second qubit, leading to the state $|\psi\rangle = \frac{1}{2}(|00\rangle + |11\rangle)$.
The sequence of gates of a program is called a quantum \emph{circuit}.

Figure~\ref{fig:ex_quantum_program_draw}, on the left, shows a pictorial representation of the program.
The figure includes two \emph{measurement} gates, shown in black, which store the result into a classical register of two bits.
Once the circuit has been defined, it is executed for some number of \emph{shots} (line~\ref{line:simulator_1024_runs} of Figure~\ref{fig:ex_quantum_program_code}) to account for the probabilistic nature of quantum programs.
The execution produces a \emph{distribution of output bit-strings}, shown on the right of Figure~\ref{fig:ex_quantum_program_draw}.
For the example program, the only two outcomes possible are bit-strings with either both 0 or both 1.

The ability to describe and execute quantum programs is provided by a quantum computing platform.
The above example is based on IBM's popular~\cite{wangQDiffDifferentialTesting2021, mendiluzeMuskitMutationAnalysis2021, zhaoBugs4QBenchmarkReal2021a} Qiskit platform~\cite{QiskitQiskit2021}, where programs are expressed using a Python API.
The platform then compiles and executes the program on a \emph{backend}, i.e., either a quantum computer or a simulator.
Part of the compilation is implemented in a \emph{transpiler}, which optimizes the circuit and prepares it for the backend.
Because different quantum computers offer different hardware gates, called the \emph{gate set}, the platform translates the program to the available sequences of gates.
How a program gets mapped to hardware is also influenced by the physical connections between qubits, which are represented in the so-called \textit{coupling map} in Qiskit.

\section{Approach}

The following presents the \methodName{} approach for metamorphic testing of quantum computing platforms.
We start with an overview and the overall algorithm (Section~\ref{sec:overview}), followed by the three main steps: generating programs (Section~\ref{sec:program_generation}), applying metamorphic transformations (Section~\ref{sec:metamorphic_framework}), and comparing the behavior of program executions (Section~\ref{sec:behavior_comparison}).

\subsection{Overall Algorithm}
\label{sec:overview}

\begin{figure}[t]
  \centering
  \includegraphics[width=0.43\textwidth]{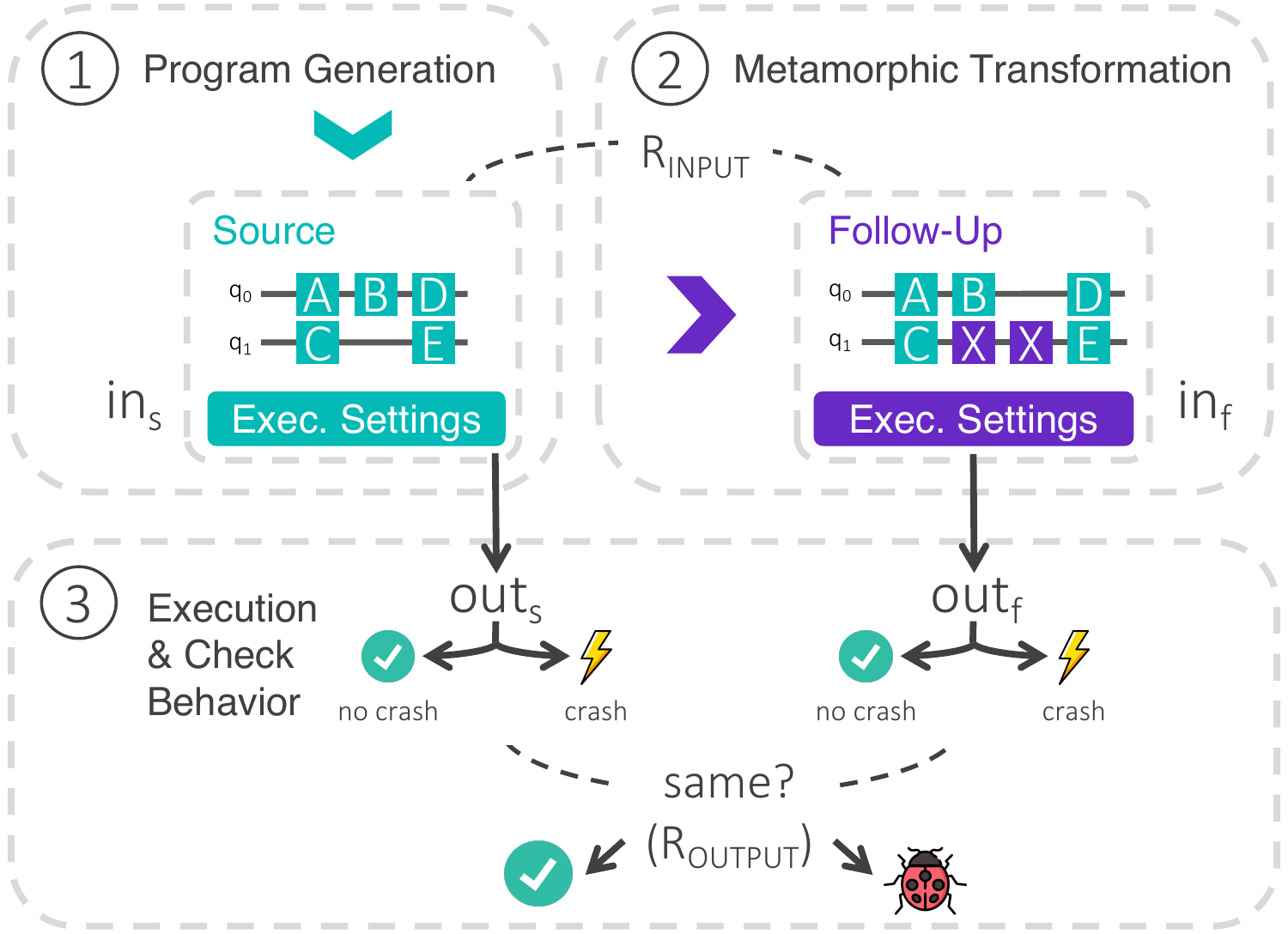}
  \caption{Overview of our approach.}
  \label{fig:overview}
\end{figure}

Figure~\ref{fig:overview} gives a high-level overview of \methodName{} and its three main steps.
At first, a program generator creates an initial quantum program, referred as the source program.
Then, by applying a sequence of metamorphic program transformations, the approach derives a follow-up program that is in a specific relationship with the source program.
Finally, the approach executes the two programs and checks whether their behaviors conform to the expected output relationship.

\begin{algorithm}[t]
  \caption{\methodName{} Approach}\label{alg:our_approach}
  \SetKwInOut{Input}{Input}
  \Input{
    Program generator $G$\\
    Metamorphic relationships $M$\\
    Behavior comparison component $C$\\
  }
  \KwResult{Likely bug-revealing pairs $B$ of programs}
  $B \gets \varnothing$\\
  \While{\normalfont{time budget} $t_{budget}$ \normalfont{not up}}{
    $\mathit{in}_{s} \gets G.generateProgram()$\Comment*[r]{STEP 1}\label{line:generate}
    $n_{toApply} \gets random(1,~max_{M})$\Comment*[r]{STEP 2}\label{line:mmStart}
    $\mathit{in}_{f} \gets \mathit{in}_{s}$\;
    \While {$n_{applied} < n_{toApply}$}{
      $m \gets sample(M)$\;
      \If{$m.checkPrecondition(\mathit{in}_{f})$}{\label{line:precond}
        $\mathit{in}_{f} \gets m.apply(\mathit{in}_{f})$\;
        $n_{applied} \gets n_{applied} + 1$\;
      }
      \If{$m$ \normalfont{is not semantics-preserving}}{
		break\;
      }
    }\label{line:mmEnd}

    $out_s, out_f \gets C.execute(\mathit{in}_{s}, \mathit{in}_{f})$\Comment*[r]{STEP 3}\label{line:compareStart}

    \If{$C.checkRelation(m_{last},out_s,out_f)$}{
      $B \gets B \cup \{(\mathit{in}_{s},~\mathit{in}_{f}) \}$\;
    }\label{line:compareEnd}
  }
  \Return $B$
\end{algorithm}

Algorithm~\ref{alg:our_approach} describes how \methodName{} composes the three main steps, represented as a program generator $G$, a set of metamorphic relationships $M$, and a component $C$ for comparing the behavior two quantum programs.
The main loop of the algorithm continuously generates and checks new pairs of source and follow-up programs until exceeding a configurable time budget.
After each iteration, both programs are discarded, making each iteration independent from the previous one and preventing \methodName{} from mutating previously crashed programs.
Finally, the algorithm returns a set $B$ of pairs of programs that expose unexpected behavior.

As the first step in the main loop of the algorithm, the program generator $G$ creates a new quantum program $\mathit{in}_s$ using a combination of template-based and grammar-based code generation (line~\ref{line:generate}).
A ``program'' here means source code that defines a quantum circuit and its execution setting, e.g.,  the type of backend to use or the transpiler's settings.
Then, the second step of the algorithm applies a sequence of transformations sampled from the metamorphic relationships $M$ to create a follow-up program $\mathit{in}_f$ (lines~\ref{line:mmStart} to~\ref{line:mmEnd}).
Each metamorphic relationship has a precondition under which its transformation may be applied.
Most of the transformations are designed to be semantics-preserving, in which case the algorithm may continue to apply further transformations.
The approach also includes two transformations that do not preserve the semantics.
Once such a transformation gets applied, the algorithm stops applying further transformations, which has the benefit that only the last transformation determines the expected output relationship.
Finally, the third step compares the behavior of the source program $\mathit{in}_s$ and the final follow-up program $\mathit{in}_f$ (lines~\ref{line:compareStart} to~\ref{line:compareEnd}).
The outcome of executing a program may be a crash or non-crashing behavior.
In the latter case, the platform repeatedly executes the circuit and summarizes the output into a distribution of bit-strings.

\subsection{Program Generation}
\label{sec:program_generation}

A naive approach to generating quantum programs might consider all elements offered by the quantum programming language, e.g., all APIs offered by Qiskit, and combine them at random.
However, such an approach would yield mostly invalid programs that crash and do not deeply test the platform.
The reason is that quantum programs need to follow a particular structure and respect various domain-specific constraints.
The program generator in \methodName{} is a combination of template-based and grammar-based code generation.
The template-based part ensures that the created programs follow the typical structure of quantum programs.
The grammar-based part is designed to cover a diverse range of possible programs by randomly combining gates with each other.
Both parts are based on concepts available across different quantum computing platforms, such as circuits, registers, gates and executing programs with a specific backend.

\begin{figure}[t]
  \begin{lstlisting}[language=Python]
  # Section: Prologue
  <ALL_IMPORTS>
  # Section: Circuit
  qr = QuantumRegister(<N_QUBITS>, name='qr')
  cr = ClassicalRegister(<N_QUBITS>, name='cr')
  qc = QuantumCircuit(qr, cr, name='qc')
  <GATE_OPS>
  # Section: Measurement
  qc.measure(qr, cr)
  # Section: Transpilation/compilation
  qc = transpile(qc,
    basis_gates=<TARGET_GATE_SET>,
    optimization_level=<OPT_LEVEL>,
    coupling_map=<COUPLING_MAP>)
  # Section: Execution
  simulator = Aer.get_backend(<BACKEND_NAME>)
  counts = execute(qc, backend=simulator,
    shots=<N_SHOTS>).result().get_counts(qc)\end{lstlisting}
  \caption{Template to generate quantum programs.}
  \label{fig:source_template}
  \end{figure}

Figure~\ref{fig:source_template} shows our template for generating quantum programs.
The placeholder \code{$\langle$ALL\_IMPORTS$\rangle$} gets replaced by the imports of all the dependencies used in a program.
In the circuit section, the template creates a quantum register and a classical register, both of size \code{$\langle$N\_QUBITS$\rangle$}, and assembles them into a quantum circuit.
The non-terminal \code{$\langle$GATE\_OPS$\rangle$} is expanded using the grammar described in Figure~\ref{fig:grammar_program_generation}, which yields a sequence of gates that act on the available qubits and bits.
Each instruction acts on a number of qubits between one and five, and contains a suitable gate that operates on them.
The indices of the target qubits are selected randomly among the integers \code{$\langle$INT$\rangle$} compatible with the maximum number \code{$\langle$N\_QUBITS$\rangle$} of qubits available.
Each gate receives a specific number of parameters, which the generator chooses among the floating point numbers \code{$\langle$FLOAT$\rangle$}.
For brevity, Figure~\ref{fig:grammar_program_generation} shows only an excerpt of the grammar.
Moving back to the template in Figure~\ref{fig:source_template}, in the transpilation section the generator replaces \code{$\langle$OPT\_LEVEL$\rangle$} with an integer from 0 to 3 indicating an optimization level, and \code{$\langle$TARGET\_GATE\_SET$\rangle$} and \code{$\langle$COUPLING\_MAP$\rangle$} with two \code{None} placeholders.
Finally, in the execution section of the program, the generator replaces \code{$\langle$BACKEND\_NAME$\rangle$} with a backend and selects the number \code{$\langle$N\_SHOTS$\rangle$} of shots to use in the execution.
For determining the right number of shots to run the program, we use a sample estimation technique proposed in prior work~\cite{wangQDiffDifferentialTesting2021}. %

\begin{figure}[t]
  \centering
  \begin{grammar}

    <GATE\_OPS> ::= <INSTR><EOL><GATE\_OPS> | <EOL>

    <INSTR> ::= <INSTR\_1Q> | <INSTR\_2Q> | ... | <INSTR\_5Q>

    <INSTR\_1Q> ::= "qc.append("<GATE\_1Q>","\\ " qregs=[qr["<INT>"]])"

    <INSTR\_2Q> ::= "qc.append("<GATE\_2Q>","\\ "qregs=[qr["<INT>"],qr["<INT>"]")

    <GATE\_1Q> ::= <HGate> | <RZGate> | ...

    <HGate> ::= "HGate()"

    <RZGate> ::= "RZGate("<FLOAT>")"

    <GATE\_2Q> ::= <CXGate> | <CRZGate> | ...

    <CXGate> ::= "CXGate()"

    <CRZGate> ::= "CRZGate("<FLOAT>")"

    <EOL> ::= "\\n"
  \end{grammar}
  \caption{Subset of the grammar to generate a sequence of gate operations.}
  \label{fig:grammar_program_generation}
\end{figure}

Our implementation of \methodName{} targets the Qiskit platform, which is highly popular and has been studied also by previous work~\cite{zhaoIdentifyingBugPatterns2021a, zhaoBugs4QBenchmarkReal2021a,paltenghiBugsQuantumComputing2022, fingerhuthOpenSourceSoftware2018}, but we believe our approach could be easily extended to other quantum computing platforms.
The generator supports a total of 45 gates, i.e., all but three gates expressible in Qiskit.
The 45 supported gates have up to four parameters and can act on up to five qubits.
The missing three gates are excluded due to deprecation, presence of non-float parameters, and missing documentation.
We limit the generation to a maximum of 30 consecutive gates to keep the execution time of programs within reasonable limits.

\subsection{Metamorphic Testing Framework}
\label{sec:metamorphic_framework}

A key technical contribution of \methodName{} is a set of ten metamorphic relationships.
We classify their corresponding transformations into three categories:
\textit{circuit transformations}, which modify the circuit;
\textit{representation transformations}, which change the intermediate representation used to represent the circuit, and
\textit{execution transformations}, which affect the execution environment.%
Table~\ref{tab:metamorphic_relationships} summarizes all transformations.
Some of them have a precondition, checked at line~\ref{line:precond} of Algorithm~\ref{alg:our_approach}, which ensures that the resulting follow-up program is indeed expected to result in behavior described by the output relationship.
All transformations in Table~\ref{tab:metamorphic_relationships}, except for \textit{Change qubit order} and \textit{Partitioned execution}, are semantics-preserving, and hence, the expected output relationship is equivalence, i.e., the source program and the follow-up program are expected to behave the same.
In particular, this output relationships means that \methodName{} reports a warning when the source program runs without crashing but the follow-up program produces a crash.

\begin{table}[t]
  \centering
  \caption{Metamorphic relationships and their preconditions.}
  \begin{tabular}{@{}p{5.5em}p{11em}p{12em}@{}}
  \toprule
  Category & Name & Precondition \\
  \midrule
  \multirowcell{2}[0ex][l]{Circuit \\transformation} & Change of qubit order & - \\
  & Inject null-effect operation & -\\
  & Add quantum register & Coupling map not fixed \\
  & Inject parameters & - \\
  & Partitioned execution & Non-interacting subsets of qubits \\
  \midrule
  \multirowcell{2}[0ex][l]{Representation \\transformation}  & Roundtrip conversion via QASM & - \\
  \midrule
  \multirowcell{2}[0ex][l]{Execution \\transformation} & Change of coupling map & No added register \\
  & Change of gate set & -\\
  & Change of optimiz.\ level & -\\
  & Change of backend & -\\
  \bottomrule
  \end{tabular}
  \label{tab:metamorphic_relationships}
\end{table}

\subsubsection{Circuit Transformations}

These transformations exploit the properties of the gate model of computation, such as the entanglement of qubits, the presence of registers and the properties of reversible computing.

\paragraph{Change of qubit order}
Inspired by the bug pattern ``incorrect qubit order''~\cite{paltenghiBugsQuantumComputing2022}, this transformation changes the order of qubits in the quantum register.
Specifically, the transformation maps the qubit indices to new positions and then adapts the gates accordingly.
Referring to the grammar in Figure~\ref{fig:grammar_program_generation}, the transformation applies a bijective mapping between the \code{$\langle$INT$\rangle$} values of the source and follow-up programs.

\begin{figure*}[t]
  \begin{subfigure}[t]{.3\linewidth}
    \includegraphics[width=\linewidth]{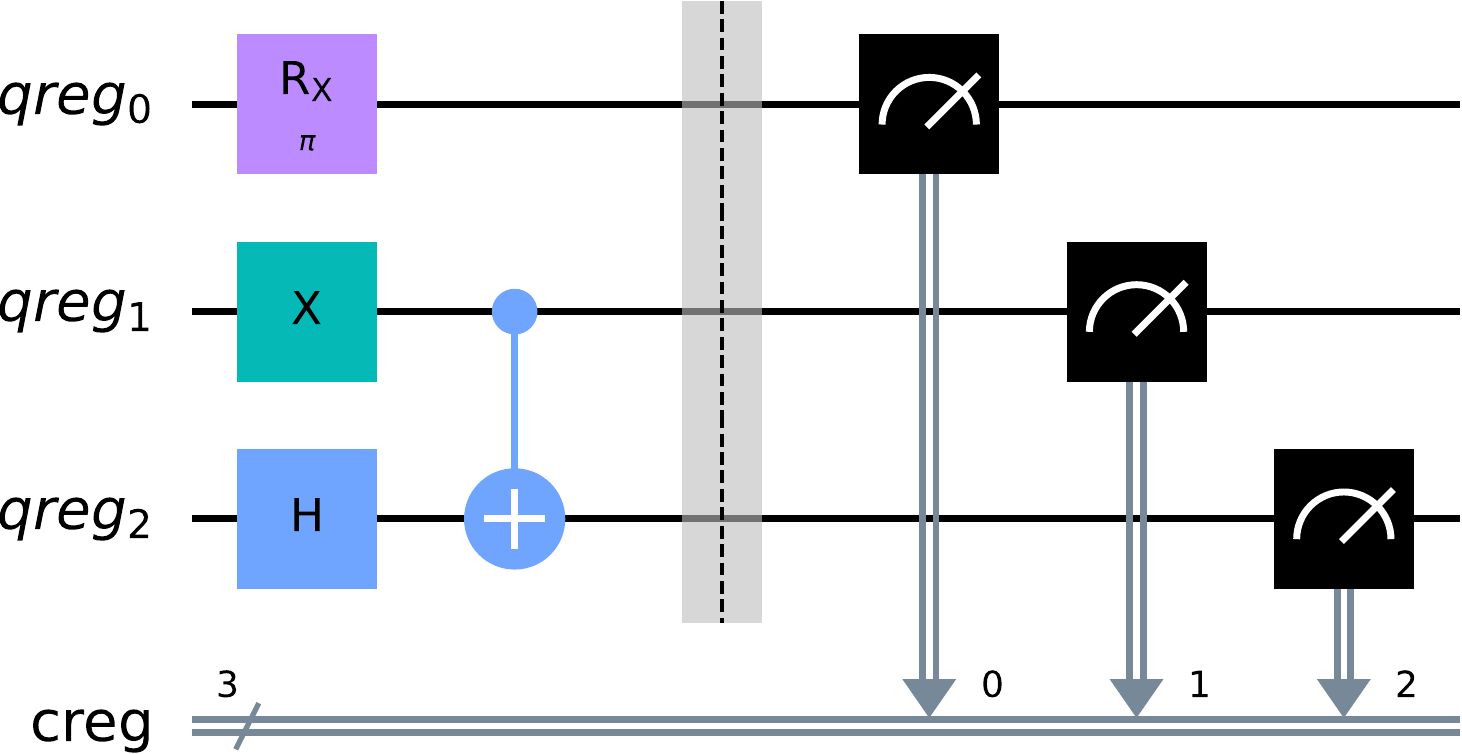}
	\caption{Source program.}
    \label{fig:source_prototypical}
  \end{subfigure}
  \hfill
  \begin{subfigure}[t]{.3\linewidth}
    \includegraphics[width=\linewidth]{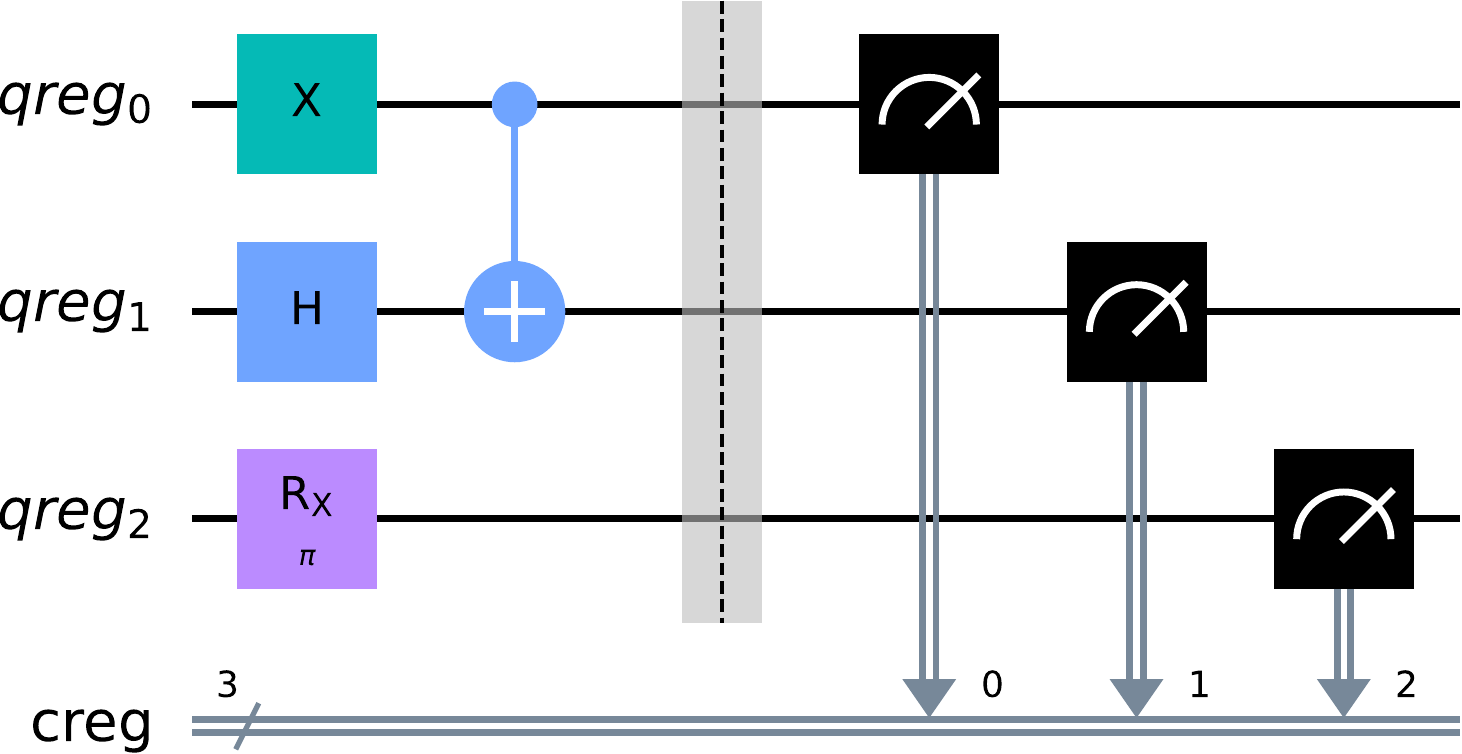}
	\caption{Follow-up program after ``change of qubit order''.}
	\label{fig:follow_change_qubit_order}
  \end{subfigure}
  \hfill
  \begin{subfigure}[t]{.23\linewidth}
    \includegraphics[width=\linewidth]{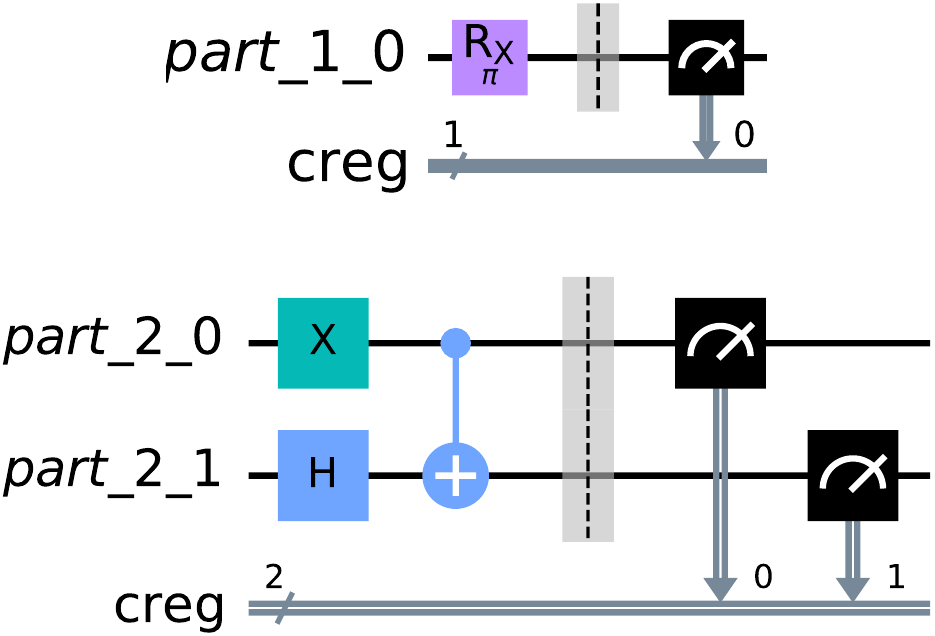}
    \caption{Follow-up program after ``Partitioned Execution''.}
    \label{fig:follow_partitions}
  \end{subfigure}
  \caption{Examples of metamorphic transformations.}
\end{figure*}

For example, consider the source circuit of Figure~\ref{fig:source_prototypical}, which has a two-qubit gate operating on qubits~1 and~2.
Applying the transformation with the qubit mapping $m = \{0\rightarrow 2; 1\rightarrow 0; 2 \rightarrow 1\}$ results in Figure~\ref{fig:follow_change_qubit_order}, where the two-qubit gate now operates on qubits~0 and~1.
The final measurement gates are not affected by the qubit mapping.
Instead, the approach applies a function to all the output bit-strings of the follow-up program that applies the inverse of $m$ to the order of measured qubits.
In the example, suppose we obtain an output bit-string \code{001} by the follow-up program.
The approach will turn it into a bit-string \code{100}, because index~2 in the follow-up program corresponds to index~0 in the source program.
After re-mapping the measurements, the two resulting output distributions are expected to be equivalent.

\paragraph{Inject null-effect operations}
In quantum computing, any operation or gate, with the exception of the measurement gate, never looses any information, and hence, can be reverted by a suitable inverse operation.
This metamorphic transformation exploits reverse computations by inserting into the main circuit a sub-circuit that performs a sequence of gate operations followed by its inverse, so that the overall effect is null.
Referring to the grammar in Figure~\ref{fig:grammar_program_generation}, the transformation injects new code between the gates generated by \code{$\langle$GATE\_OPS$\rangle$}.
The inserted sub-circuit may include an arbitrary number of gates and act on an arbitrary number of available qubits.
The only restriction is that no measurement is introduced, because otherwise it would destroy the quantum state and change the result with respect to the source program.

Figure~\ref{fig:code_null_effect_injection} gives an example of injected code.
The inverse is produced via a function called \code{inverse} (line~\ref{line:inverse_funct_call}), which is offered by most quantum computing platforms and reverses the effect of a sub-circuit.

\begin{figure}[t]
  \begin{lstlisting}[language=Python, escapechar=\%]
subcirc = QuantumCircuit(qr, cr, name='subcirc')
subcirc.append(RXGate(6.12), qargs=[qr[0]], cargs=[])
# ... sequence of additional gates
qc.append(subcirc, qargs=qr, cargs=cr)
qc.append(subcirc.inverse(), qargs=qr, cargs=cr)%\label{line:inverse_funct_call}%\end{lstlisting}
  \caption{Example of code inserted by the ``inject null-effect operation'' transformation.}
  \label{fig:code_null_effect_injection}
\end{figure}

\paragraph{Add quantum register}
Enlarging the set of available qubits by adding a new and unused quantum register should not affect the computation on the existing qubits.
This transformation exploits this property by randomly adding new quantum registers to the circuit of the follow-up program.
Referring to Figure~\ref{fig:source_template}, the new register is added right before or after the measurement section.
This transformation cannot be performed when the coupling map has been specified before via the \textit{Change of coupling map} transformation, since the addition of a register would make the coupling map too small.

\paragraph{Inject parameters}
Given the recent interest in quantum machine learning, quantum computing platforms offer abstractions to support the parametrization of quantum circuits~\cite{schuldIntroductionQuantumMachine2015}.
One of the subfields of quantum machine learning aims to use quantum circuits and the parameters of their gates as a quantum version of artificial neural networks~\cite{schuldQuestQuantumNeural2014}.
This transformation creates such parameterized circuits by replacing one or more floating point literals \code{$\langle$FLOAT$\rangle$} in the source program with a corresponding \code{Parameter('a')} object.
Then, before the transpilation stage, the transformation binds all the free parameters to the original literal values.
In analogy to traditional programs, this transformation resembles moving a literal value into a variable.

\paragraph{Partitioned execution}
Some source programs have two subsets of qubits such that there is no gate operation that involves qubits from both subsets.
In this case, the source program performs two independent computations that can be executed in parallel.
This transformation separates the circuit of such programs into two sub-circuits, executes them individually, and then post-processes the results to derive the distribution of the overall program.
The output distribution of the source program has bit-strings of size \code{$\langle$N\_QUBITS$\rangle$}, whereas the result of the follow-up program consists of two distributions with bit-strings of sizes $a$ and $b$, where \code{$\langle$N\_QUBITS$\rangle$}$= a + b$.
To reconstruct an output distribution of size \code{$\langle$N\_QUBITS$\rangle$} also for the follow-up program, the approach computes the Cartesian product of the output distributions of the two sub-circuits:
$U_s|\phi\rangle = U_{f1}|\phi\rangle_{1} \otimes U_{f2}|\phi\rangle_{2}$,
where $U_s$ represents the gates of the source program and $|\phi\rangle$ represents all qubits, $U_{f1}$ and $U_{f2}$ correspond to the two sub-circuits, and $|\phi\rangle_{1}$ and $|\phi\rangle_{2}$ are the two subsets of qubits.

Figure~\ref{fig:follow_partitions} shows two partitions derived from the circuit in Figure~\ref{fig:source_prototypical}, the first partition with a single qubit and an ``rx'' gate, and the second with two qubits and the remaining gates.

\subsubsection{Representation Transformations}

The following is a transformation that acts on the representation of the quantum program, without affecting its computation or execution environment.

\paragraph{Roundtrip conversion via QASM}
OpenQASM~\cite{crossOpenQuantumAssembly2017}, or short QASM, is the de-facto standard assembly language for quantum programs.
Many quantum computing platforms offer API calls to convert to and from it, and virtually all circuits can be expressed in the QASM format.
Because correctly converting to and from QASM is an important prerequisite for the interoperability of quantum computing platforms, \methodName{} comes with a transformation that exercises these parts of the platform under test.
The transformation converts the quantum circuit to the QASM format and then parses the QASM code again to reconstruct the original circuit.
To implement the roundtrip conversion in Qiskit, the transformation uses these API calls: \code{qc = qc.from\_qasm\_str(qc.qasm())}.
The approach performs this transformation right before the execution section in Figure~\ref{fig:source_template}.

\subsubsection{Execution Transformations}

The third category of transformations is about adapting the execution environment.
Given the currently available quantum hardware, called ``noisy intermediate-scale quantum'' (NISQ) devices~\cite{preskillQuantumComputingNISQ2018}, executing many generated programs on quantum hardware results in noise-induced behavioral differences~\cite{wangQDiffDifferentialTesting2021}.
To avoid false positives caused by hardware limitations, while still being able to find bugs in the software stack of quantum computing, \methodName{} focuses on executing programs on simulators.

\paragraph{Change of coupling map}
This transformation replaces the placeholder \code{$\langle$COUPLING\_MAP$\rangle$} in the program template with a randomly created coupling map.
The coupling map describes the physical connections between qubits as list of pairs of qubit indices.
\methodName{} ensures the coupling map to yield a connected graph of qubits so that no qubit is isolated.
An example of a coupling map for our program in Figure~\ref{fig:source_prototypical} is \code{[[0,1],[1,2]]}.

\paragraph{Change of gate set}
During transpilation, a given quantum program is converted to be compatible with a specific target device, which often involves translating the gates to the natively supported gates.
This transformation exercise this translation step by replacing the \code{$\langle$TARGET\_GATES$\rangle$} in the program template with a universal gate set, such as the \code{["rx", "ry", "rz", "p", "cx"]} gates~\cite{williamsQuantumGates2011}.
\methodName{} currently supports three universal gate sets but could be easily extended.

\paragraph{Change of optimization level}
The final two transformations are inspired by work on compiler testing~\cite{chenEmpiricalComparisonCompiler2016}.
One transformation replaces the \code{$\langle$OPT\_LEVEL$\rangle$} in the program template with another level between 0 and 3, which is not expected to affect the final output of a program.

\paragraph{Change of backend}
This transformation replaces the non-terminal \code{$\langle$BACKEND\_ NAME$\rangle$} in the program template with another available backend.
Different simulators typically have completely different implementations.
A single simulator often offers two variants, running on a CPU and GPU respectively, which we treat as two separate backends.
In total, \methodName{} supports eight different backends.

\subsection{Comparing Execution Behavior}
\label{sec:behavior_comparison}

The third and final step of \methodName{} is to execute both the source program and the follow-up program.
If the two programs expose different behaviors, \methodName{} adds them to the set of likely bug-revealing pairs of programs.

We perform this comparison at two levels.
The first level identifies cases where one program runs without any crash, but the other program crashes, called a \emph{crash difference}.
Our program generator (Section~\ref{sec:program_generation}) is designed to create source programs that do not crash.
However, applying the metamorphic transformations may trigger some bugs in the tested platform that manifest through a crash.

The second level compares the measured output bits of two non-crashing programs.
Due to the probabilistic nature of quantum programming, precisely comparing the output bit-strings would be misleading.
Instead, \methodName{} repeatedly executes each circuit for the specified number of shots and then compares the two output distributions.
We use the Kolmogorov-Smirnov test~\cite{kolmogorovSullaDeterminazioneEmpirica1933, smirnovTableEstimatingGoodness1948} to assess the statistical significance of the difference between the two distributions, as done in previous work~\cite{wangQDiffDifferentialTesting2021}.
\methodName{} reports any pair of programs with a p-value below $5\%$ as a statistically significant \emph{distribution difference}.

\section{Implementation}

\methodName{} is implemented in Python and tested on the latest Qiskit $0.19.1$ version at the time of performing the evaluation.
The implementation is designed in a modular way with four main components: (1) the \methodName{} core, which is responsible for the orchestration of the various steps of the approach, (2) a program generator, which produces valid programs according to the API of the platform, (3) an extensible set of metamorphic transformations, which apply lightweight program transformations based on the API of the platform, (4) a component that spots any differences in execution behavior.
\methodName{} currently supports Qiskit as a first target platform, but could be extended to other quantum computing platforms.

\section{Evaluation}

Our evaluation focuses on the following research questions:

\begin{itemize}
  \item RQ1: How many warnings does \methodName{} produce?
  \item RQ2: What real-world bugs does \methodName{} find in Qiskit?
  \item RQ3: How does \methodName{} compare to prior work on testing quantum computing platforms~\cite{wangQDiffDifferentialTesting2021}.
  \item RQ4: To what extent do the different metamorphic relations contribute to the warnings and bugs found?
  \item RQ5: How efficient is \methodName{} and what are the most time-consuming components?
\end{itemize}

All experiments are run on a machine with 48 CPU cores (Intel Xeon Silver, 2.20GHz), two NVIDIA Tesla T4 GPUs with 16GB memory each, and 252GB of RAM, which is running Ubuntu 18.04.5.

\subsection{RQ1: Warnings Produced by \methodName{}}

\begin{table}[t]
  \centering
  \caption{Warnings produced in 48 hours by the MorphQ approach and using only QDiff's transformations~\cite{wangQDiffDifferentialTesting2021}.}
  \begin{tabular}{@{}lrr|rr@{}}
  \toprule
  & \multicolumn{2}{c}{MorphQ} & \multicolumn{2}{c}{QDiff Transf.} \\
  \cmidrule{2-3}
  \cmidrule{4-5}
  & No. & \%  & No. & \% \\
  \midrule
  Tested program pairs& 8,360& 100.0 & 51,271& 100.0 \\
  \hspace{1em} $\hookrightarrow$ Crashes in source program & 0 & 0.0 & 0 & 0.0\\
  \hspace{1em} $\hookrightarrow$ Crashes in follow-up program & 1,943 & 23.2 & 0 & 0.0\\
  \hspace{1em} $\hookrightarrow$ Successful executions & 6,417 & 76.8 & 51,271 & 100.0\\

  \hspace{3em} $\hookrightarrow$ Distribution differences & 56 & 0.7 & 528 & 1.0\\

\bottomrule
  \end{tabular}
  \label{tab:warnings_distribution}
\end{table}

\begin{figure}[t]
  \centering
  \includegraphics[width=0.42\textwidth]{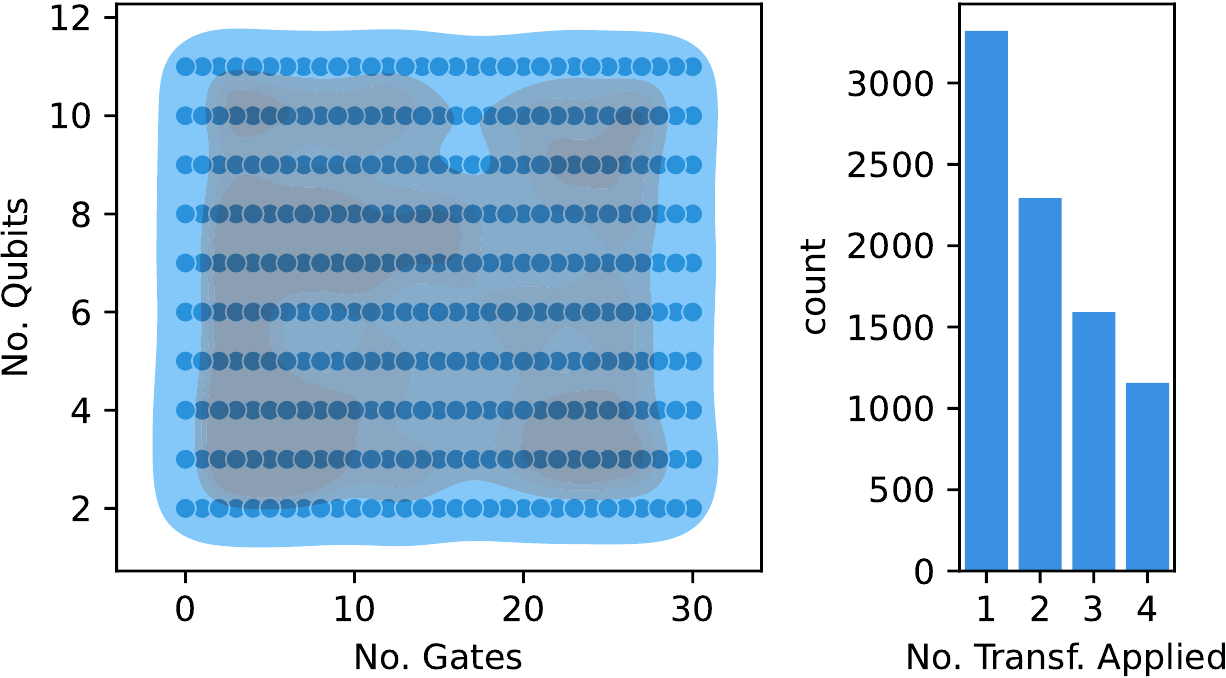}
  \caption{Characteristics of the programs generated by MorphQ.}
  \label{fig:dataset_stats}
\end{figure}

This research question quantitatively evaluates \methodName{}'s effectiveness at finding unexpected behavior.
We run the approach for a total of 48 hours, as done in previous work~\cite{wangQDiffDifferentialTesting2021}, and summarize the results in Table~\ref{tab:warnings_distribution}.
Over this period, the program generator produces a total of \totalTestedProgramPairs{} programs.
In Figure~\ref{fig:dataset_stats}, on the left, we report the distribution of the number of qubits and number of gates in the generated programs, where a darker color means a higher density of programs.
The right side of the figure shows how many follow-up programs are generated by applying a specific number of transformations.

All programs generated by \methodName{} execute without crashing, which confirms that our template-based and grammar-based generation technique is successful at generating valid quantum programs.
Applying metamorphic relations to these programs leads to a program crash in \percCrashesBenchmark{} of the cases, and hence, is reported as a crash difference.
Out of the non-crashing executions, a small percentage of a total of \totalPairsWithDistrDifference{} programs exposes a distribution difference.

\begin{answerbox}
\textbf{Answer to RQ1}: The program generation successfully creates only valid quantum programs, and \methodName{} is effective in producing numerous warnings, e.g., by inducing \percCrashesBenchmark{} of all follow-up programs to crash.
\end{answerbox}

\subsection{RQ2: Real-World Bugs Found}

To evaluate \methodName{}'s ability to find real-world bugs, we inspect a sample of warnings produced over a period of about \nDaysRun{} days.

\subsubsection{Crash Differences}

\begin{figure}[t]
  \centering
  \includegraphics[width=0.43\textwidth]{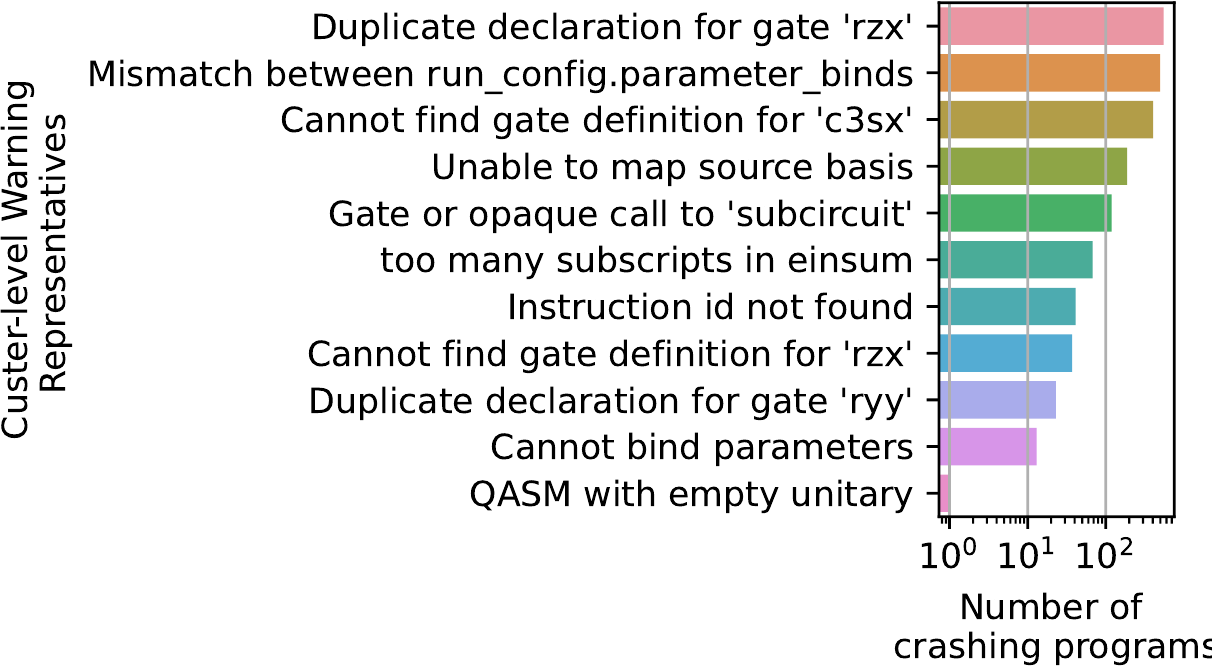}
  \caption{Regex-based manual clustering of warnings.}
  \label{fig:clustering}
\end{figure}

Because crash-inducing bugs are the most critical, as they impede developers from running their programs at all, we focus most of our attention on them.
Before inspecting program pairs with a crash difference, we semi-automatically cluster the warnings based on their crash message.
To this end, we abstract program-specific references, such as line numbers, variable names, and file names, and then assign all warnings with the same abstracted message into a cluster.
For example, ``Duplicate declaration for gate 'ryy', line 4, fileA'' and ``Duplicate declaration for gate 'ryy', line 5, fileB'' are assigned to the same cluster.
Figure~\ref{fig:clustering} shows the resulting clustering of warnings.

We then randomly select a few failing follow-up programs from each cluster for manual inspection.
The inspection procedure consists in manually reversing each transformation in the follow-up program, one at the time, until we find which transformation is responsible for the crash.

Then, once detected which transformation or combination of transformations is responsible, we reduce the gate operations in the program in a delta debugging~\cite{zellerIsolatingCauseeffectChains2002}-like manner until we identify the minimal sequence of operations to trigger the crash.
This manual process requires about 15 minutes per program, and it is feasible since the programs have at most 30 operations and four transformations.
Further automating the crash clustering and the minimization is left for future work.

Table~\ref{tab:real_bugs} summarizes the results of our manual inspection.
For each warning, we report the reference to the bug report\footnote{Removed for double-blind review. See supplementary material for anonymized versions of the bug reports.}, its status, whether it was a new or duplicated bug report, the crash message, and what metamorphic transformation(s) are required to trigger the bug.
Over the course of this study, we have filed a total of \nBugReports{} bug reports in the Qiskit repository.
So far, \nBugsFound{} of the reports have been confirmed by the developers as bugs.
The following describes some representative examples of the inspected warnings.

\newcommand{\IDQargsNotInThisCircuit}{1}
\newcommand{\IDTooManySubscriptsInEinsumNumpy}{2}
\newcommand{\IDGateOrOpaqueCallToSubcircuit}{3}
\newcommand{\IDDuplicateDeclarationForGateRzx}{4}
\newcommand{\IDInstructionIdNotFound}{5}
\newcommand{\IDMismatchBetweenParameterbinds}{6}
\newcommand{\IDCannotFindGateDefinitionForCsx}{7}
\newcommand{\IDCannotBindParametersNotPresentInTheCircuit}{8}
\newcommand{\IDQasmGateDefinitionWithNoOperands}{9}
\newcommand{\IDCannotFindGateDefinitionForRzx}{10}
\newcommand{\IDDuplicateDeclarationForGateRyy}{11}
\newcommand{\IDCannotFindGateDefinitionForUnitary}{12}
\newcommand{\IDCannotFindGateDefinitionForRcccx}{13}

\begin{table*}[t]
  \centering
  \caption{Real-world bugs and warnings found by \methodName{}.}

  \begin{tabular}{@{}rlllp{5cm}p{7.3cm}@{}}
  \toprule
   ID & Report &    Status &   Novelty &                                     Crash message &                                                                Metamorphic transformations \\
  \midrule
    1 &  \#7694 & confirmed &       new &                         qargs not in this circuit &                                       Change of optimization level, Change of coupling map \\
    2 &  \#7700 & confirmed &       new &             too many subscripts in einsum (numpy) &                                Change of optimization level, Inject null-effect operations \\
    3 &  \#7750 & confirmed &       new &               Gate or opaque call to `subcircuit' &                               Roundtrip conversion via QASM, Inject null-effect operations \\
    4 &  \#7749 & confirmed & duplicate &              Duplicate declaration for gate `rzx' &                                                              Roundtrip conversion via QASM \\
    5 &  \#7641 & confirmed & duplicate &                          Instruction id not found &                                                                         Change of gate set \\
    6 &  \#7326 & confirmed & duplicate &                  Mismatch between parameter\_binds &                                                                          Inject parameters \\
    7 &  \#7756 & confirmed & duplicate &            Cannot find gate definition for `c3sx' &                                                              Roundtrip conversion via QASM \\
    8 &  \#7748 &     fixed &       new & Cannot bind parameters not present in the circuit &                                                                          Inject parameters \\
    9 &  \#8224 &     fixed &       new &             QASM gate definition with no operands & Change of optimization level, Roundtrip conversion via QASM, Inject null-effect operations \\
   10 &  \#7769 &  reported &         - &             Cannot find gate definition for `rzx' &                               Roundtrip conversion via QASM, Inject null-effect operations \\
   11 &  \#7771 &  reported &         - &              Duplicate declaration for gate `ryy' &                               Roundtrip conversion via QASM, Inject null-effect operations \\
   12 &  \#7772 &  reported &         - &           Cannot find gate definition for unitary & Change of optimization level, Roundtrip conversion via QASM, Inject null-effect operations \\
   13 &  \#7773 &  reported &         - &           Cannot find gate definition for `rcccx' &                               Roundtrip conversion via QASM, Inject null-effect operations \\
  \bottomrule
  \end{tabular}
    \label{tab:real_bugs}
\end{table*}

\paragraph{Confirmed Bugs}

Bug~\IDTooManySubscriptsInEinsumNumpy{} is detected thanks to two different metamorphic transformations applied simultaneously, showing the importance of combining multiple transformations.
The transformations involved are: \textit{change of optimization level} and \textit{inject null-effect operations}.
Figure~\ref{fig:minimized_numpy_bug} shows the minimized follow-up program consisting of a main circuit with eleven qubits, a subcircuit with ten qubits, and an optimization pass of level 2.
This program triggers a generic Numpy error message.
As confirmed by a Qiskit developer, the bug is in a specific analysis part of the optimization, called the \code{CommutationAnalysis}.
The goal of this analysis is to find operation nodes that can commute in the direct acyclic graph representing the program.
The problem is that the implementation of this analysis relies on matrix multiplications with $n\_qubits \times 3$ dimensions, which in the case of eleven qubits is $33$, whereas the maximum dimension supported by Numpy is $32$ (\code{numpy.MAXDIM}).

\begin{figure}[t]
  \begin{lstlisting}[language=Python]
qr = QuantumRegister(11, name='qr')
cr = ClassicalRegister(11, name='cr')
qc = QuantumCircuit(qr, cr, name='qc')
subcircuit = QuantumCircuit(qr, cr, name='subcirc')
subcircuit.x(3)
qc.append(subcircuit, qargs=qr, cargs=cr)
qc.x(3)
qc = transpile(qc, optimization_level=2)
# ValueError: too many subscripts in einsum\end{lstlisting}
  \caption{Minimal follow-up program to trigger Bug~\IDTooManySubscriptsInEinsumNumpy{}.}
  \label{fig:minimized_numpy_bug}
  \end{figure}

Bug~\IDInstructionIdNotFound{} is discovered by the transformation \textit{Change of gate set}.
Whenever the transpiler has to convert a circuit that, among the other gates, includes an identity gate, then the transpiler fails.
The reason is that the identity gate is treated as a delay by the scheduler, since an identity gate operation is equivalent to a no-operation.
As a consequence, there is no translation rule for the identity gate which leads to an exception in the translation process.
The developers confirmed the bug, which had already been detected independently, and proposed a patch to fix it.

Bug~\IDGateOrOpaqueCallToSubcircuit{} is triggered by a combination of two transformations: \textit{Roundtrip conversion via QASM} and \textit{Inject null-effect operations}.
Figure~\ref{fig:minimized_subcirc_classical_bits} shows a minimized circuit that triggers the bug.
It contains a subcircuit with a classical register, which is then converted to QASM and back to a quantum circuit.
Running this code makes the QASM importer call to \code{qasm\_from\_str} produce an error caused by parsing invalid QASM code.
The root cause of the error is actually in the QASM exporter, which produces the faulty QASM code shown in Figure~\ref{fig:qasm_subcirc_classical_bits}.
A Qiskit developer confirmed this bug by saying it should have been rejected by the exporter, since it is not possible to represent sub-circuits with classical registers in QASM.

\begin{figure}[t]
\begin{lstlisting}[language=Python]
qr = QuantumRegister(2, name='qr')
cr = ClassicalRegister(2, name='cr')
qc = QuantumCircuit(qr, cr, name='qc')
subcircuit = QuantumCircuit(qr, cr, name='subcirc')
subcircuit.x(qr[0])
qc.append(subcircuit, qargs=qr, cargs=cr)
qc = QuantumCircuit.from_qasm_str(qc.qasm())
# QasmError: 'subcirc' uses 4 qubits but is declared for 2 qubits\end{lstlisting}
\caption{Minimal follow-up program to trigger Bug~\IDGateOrOpaqueCallToSubcircuit{}.}
\label{fig:minimized_subcirc_classical_bits}
\end{figure}

\begin{figure}[t]
  \begin{lstlisting}OPENQASM 2.0;
include "qelib1.inc";
gate subcircuit q0,q1 { x q0; }
qreg qr[2];
creg cr[2];
subcircuit qr[0],qr[1],cr[0],cr[1];\end{lstlisting}
  \caption{Wrong QASM code produced because of Bug~\IDGateOrOpaqueCallToSubcircuit{}.}
  \label{fig:qasm_subcirc_classical_bits}
\end{figure}

\paragraph{False Positives}
\label{sec:eval fp}

Beyond actual bugs, \methodName{} may also produce false positive warnings because the assumptions of our metamorphic relations do not hold.
We are aware of one such invalid assumption, which happens during the \textit{Change of gate set} transformation.
The transformation assumes that any circuit can be transformed into an equivalent circuit that uses only gates inside one of the universal gate sets.
While this assumption holds in theory, the implementation in Qiskit uses the A* algorithm to find an equivalent sequence of gates because exploring all possible sequences is impractical.
Because this search may fail in the computational budget provided by Qiskit, the follow-up program sometimes crashes with a ``Unable to map source basis to target basis'' crash message, which does not point to a bug in the platform, but simply a limitation of its implementation.

\subsubsection{Distribution Differences}
\label{sec:inspection_distrib_diff_morphq}

Besides crash differences, \methodName{} also warns about differences between the probability distributions that result from measurements in an initial program and a follow-up program.
As manually inspecting differences and understanding their root cause involves significant human effort, we sample and inspect ten program pairs reported to have distribution differences.
Unfortunately, all the differences turn out to be benign.
In particular, re-running the programs to see if the divergence is due to randomness or is reproducible across runs shows the differences to be a result of randomness.

A closer look at the number of program pairs with distribution differences, e.g., in Table~\ref{tab:warnings_distribution}, shows that this number is within the range of expected false positives.
When statistically identifying distribution differences, \methodName{} uses a $5\%$ threshold on the p-value (Section~\ref{sec:behavior_comparison}).
That is, observing a false positive distribution difference for up to $5\%$ of the program pairs is expected.
An effective way to identify distribution differences that are likely true positives will be interesting future work, which then can be easily plugged into \methodName{}.

\begin{answerbox}
  \textbf{Answer to RQ2}: \methodName{} has discovered \nBugReports{} bugs in the latest version of Qiskit, \nBugsFound{} of which have already been confirmed by the developers.
\end{answerbox}

\subsection{RQ3: Comparison with Prior Work}
\label{sec:eval qdiff}

\subsubsection{Bugs Found}
We compare with QDiff~\cite{wangQDiffDifferentialTesting2021}, which is the only other automated technique for testing quantum computing platforms that we are aware of.
As one way of comparing the two approaches, we compare the bugs found by \methodName{} and those reported in the QDiff paper.
During its evaluation on Qiskit, QDiff has reported distribution differences due to hardware characteristics, but no software bugs in Qiskit.
In contrast, \methodName{} discovers several software bugs in Qiskit (Table~\ref{tab:real_bugs}), none of which have been found by QDiff.

\subsubsection{Qiskit's Transformations Re-implemented in \methodName{}}
As another way of comparing with QDiff, we re-implement in the \methodName{} framework the seven semantics-preserving code transformations that QDiff uses to create test programs.
These transformations insert, delete, or change individual gates in a program.
\methodName{} applies these transformations to initial programs created by our program generator, followed by a transformation that changes the execution environment by either changing the backend or the optimization level.
The rationale for changing the execution environment is to mimic the differential testing performed by QDiff.
We then perform the same experiment as in RQ1, i.e., let the \methodName{} framework, with only the QDiff transformations, run for 48 hours.

The right block in Table~\ref{tab:warnings_distribution} shows the warnings reported in this experiment.
Unfortunately, the approach does not reveal any crashes, but only distribution differences, which matches the results reported in the QDiff paper.
On the upside, using only QDiff's transformations causes our framework to generate more follow-up programs (51,271 vs.\ 8,360).
The reason is that the follow-up programs produced by \methodName{} have longer execution times.

\subsubsection{Distribution Differences}

Since QDiff is specifically targeting distribution differences, we also inspect ten reported distribution differences as done for \methodName{} in Section~\ref{sec:inspection_distrib_diff_morphq}.
Unfortunately, performing additional re-runs of the programs that expose distribution differences causes the divergence to disappear, i.e., all the differences reported by
QDiff turn out to be benign.
Similar to the discussion in Section~\ref{sec:inspection_distrib_diff_morphq}, the results match the expected false positive rate of the statistical test.

\subsubsection{Coverage of Qiskit Code}
As a third way of comparing with QDiff, we measure the code coverage of the Qiskit platform when being tested (i) with \methodName{} and (ii) with \methodName{} using the reimplementation of QDiff's transformations.
We find that \methodName{} reaches higher coverage (\coverageMorphQ{}\% vs.\ \coverageQDiff{}\%) in the same testing budget of 48 hours, despite executing a lower number of programs.

\subsubsection{Diversity of Follow-up Programs}
\newcommand{\nuniqueAPICallPairsMorphQ}{977}
\newcommand{\nuniqueAPICallPairsQDiff}{259}

As the effectiveness of metamorphic testing depends on the ability to generate a diverse set of follow-up programs, we assess the diversity of these programs.
We perform this assessment both for the follow-up programs created by \methodName{} and by the QDiff transformations, based on source programs generated in the same way for both approaches.
For each generated follow-up program, we compute all pairs of consecutive API calls, and then we compute how many unique pairs there are among all the programs generated by an approach.
We ignore calls to ``append()'' since it is a ubiquitous call to append an instruction.
During our 48-hour experiment, the follow-up programs from QDiff's transformations have \nuniqueAPICallPairsQDiff{} unique API call pairs, whereas MorphQ's follow-up programs have \nuniqueAPICallPairsMorphQ{}, which shows a higher degree of diversity in the follow-up programs by \methodName{}.

\begin{answerbox}
  \textbf{Answer to RQ3}:
  Compared to prior work~\cite{wangQDiffDifferentialTesting2021},
  \methodName{} reveals previously undetected, crash-inducing bugs, achieves higher code coverage of the tested platform, and generates more diverse follow-up programs.
\end{answerbox}

\subsection{RQ4: Contribution of Metamorphic Transformations}

To better understand to what extent the different metamorphic transformations in \methodName{} contribute to its effectiveness, we check which transformations are more involved in reporting warnings and which are essential to expose the found bugs.

\subsubsection{Warnings}

\begin{figure}[t]
  \centering
  \includegraphics[width=0.43\textwidth]{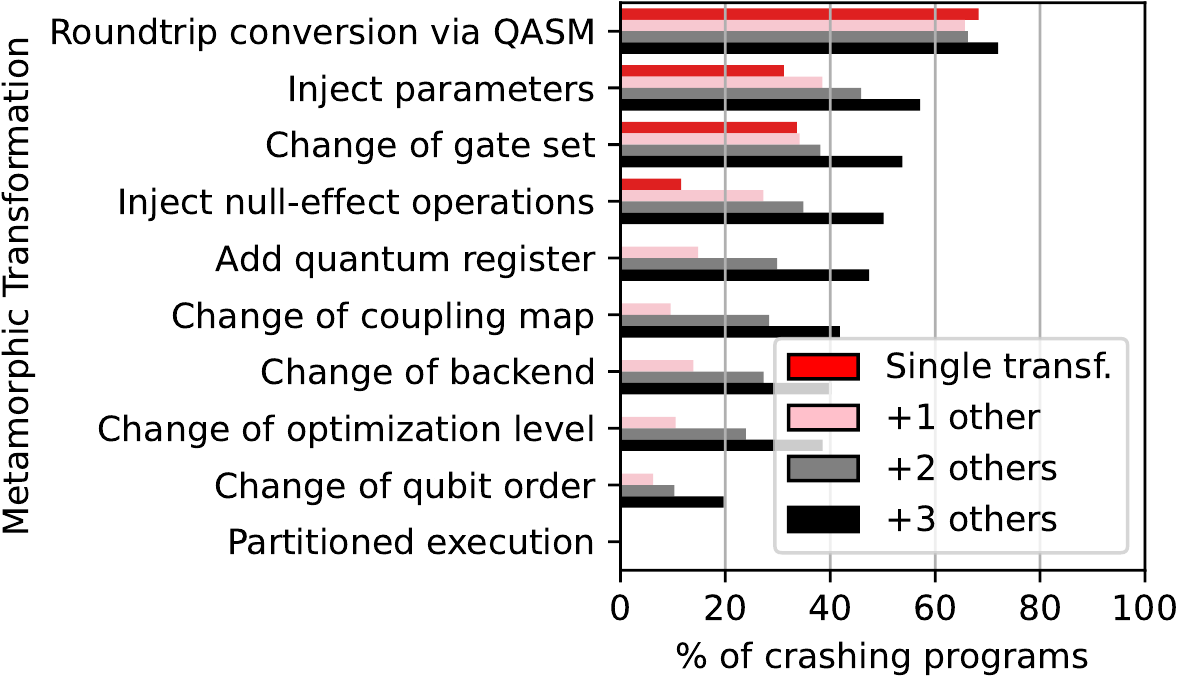}
  \caption{Percentage of crashing programs containing only a given transformation (red), the given transformation and one other transformation (pink), two others (gray), or three others (black).}
  \label{fig:contribution_of_mr_perc_crashes_vs_n_transform}
\end{figure}

Figure~\ref{fig:contribution_of_mr_perc_crashes_vs_n_transform} shows how often each transformation is involved in producing a crashing follow-up program.
Because crashes may be the result of applying one or more transformations, the figure shows the percentage of crashing programs that include only a specific transformation (red), that transformation combined with one (pink), two (gray), or three (black) others.

The transformation leading to most crashes is \textit{Change of gate set}, some of which are the false positive case discussed in Section~\ref{sec:eval fp}.
The second most commonly crash-inducing transformation is \textit{Roundtrip QASM conversion}, which shows that QASM exporter and importer is a complex, error-prone component of the platform under test.
\textit{Inject null-effect operations} and \textit{Inject parameters} also induce a sizable set of crashes, which we attribute to the fact that they exercise recently added code.
\textit{Partitioned execution} is not involved in any crashes, which we attribute both to the fact that is is applied only under a specific precondition and that it is not semantics-preserving, i.e., no other transformation gets applied afterwards.

\subsubsection{Bugs}
For each bug found by \methodName{}, we manually reduce the bug-inducing test program to keep only those metamorphic transformations that are required to expose the crash, shown in the last column of Table~\ref{tab:real_bugs}.
Finding the \nBugReports{} bugs is enabled by a total of six metamorphic transformations.
The most prevalent transformation is \textit{Roundtrip QASM conversion}.
We also find that \nBugReportsMultipleTransformations{} out of the \nBugReports{} warnings require at least two transformations, underlining the importance of combining them.

\begin{answerbox}
  \textbf{Answer to RQ4}:
  Some transformations, e.g., \textit{Roundtrip conversion via QASM} and \textit{Inject null-effect operations}, are particularly effective at revealing crashes and bugs.
  Composing multiple transformations is key to exposing \nBugReportsMultipleTransformations{} out of \nBugReports{} bugs.
\end{answerbox}

\subsection{RQ5: Time Cost per Component}
\newcommand{\avgTimeMsGenerationAndTransf}{36.9}
\newcommand{\avgTimeMsGenerationOnly}{6.2}
\newcommand{\avgTimeMsTransfOnly}{30.6}

\begin{figure}[t]
  \centering
  \includegraphics[width=0.48\textwidth]{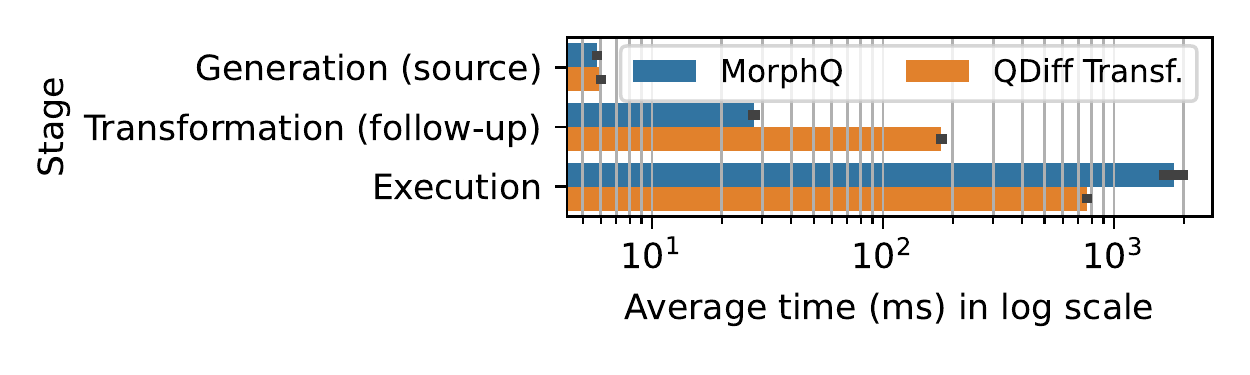}
  \caption{Time spent per component of \methodName{}.}
  \label{fig:RQ4_time_analysis}
\end{figure}

The following studies how efficient the different steps of \methodName{} are and which step takes most time.
We measure the time spent in the three main components, namely (i) generating source programs, (ii) creating follow-up programs via a series of transformations, and (iii) executing programs on simulators and compare their behavior.
Figure~\ref{fig:RQ4_time_analysis} reports the time per component, on average for a single pair of programs, during the two-day experiment from RQ1.
For comparison, we also show the results with the QDiff transformations only (see RQ3).
The by far most time-consuming step is to execute the programs, as executing larger circuits in a simulator running on classical hardware is known to be slow.
In contrast, generating and transforming programs take only \avgTimeMsGenerationOnly{}ms and \avgTimeMsTransfOnly{}ms, respectively.

\begin{answerbox}
  \textbf{Answer to RQ5}:
  Program generation and performing metamorphic transformations are efficient, together taking only \avgTimeMsGenerationAndTransf{}ms per program pair, whereas executing the programs on simulators is the most time-consuming step of the approach.
\end{answerbox}

\section{Threats to Validity}

There are some threats to the validity of our results and the conclusions to draw from them.
First, the results might be influenced by the non-deterministic, randomized nature of the program generator and the selection of transformations.
We mitigate this threat via long-running experiments, which compensate for any bias in the results one might observe with only a few generated programs.
Second, the number of warnings gives only a partial view of the effectiveness of \methodName{} due to the presence of duplicates~\cite{chenEmpiricalComparisonCompiler2016}.
To mitigate this threat we cluster warnings and inspect a sample, showing that there are at least \nBugReports{} unique bugs.
Finally, our experiments focus on a single target platform and we cannot claim that our results will generalize beyond it.
We believe the approach could also be applied to other quantum computing platforms that use a circuit-based computational model, provide similar programming abstractions, and offer QASM compatibility, such as \textit{Pytket}~\cite{sivarajahKetRetargetableCompiler2020} and \textit{Cirq}~\cite{developersCirq2021}.

\section{Related Work}

\paragraph{Quantum computing platforms}

A study by Paltenghi and Pradel~\cite{paltenghiBugsQuantumComputing2022} identifies ten quantum-specific bug patterns in quantum computing platforms, such as \textit{incorrect qubit order} and \textit{incorrect intermediate representation}, which inspired some metamorphic transformations of \methodName{}.
Other studies report how bugs in Qiskit manifest~\cite{zhaoBugs4QBenchmarkReal2021a} and discuss challenges faced by platform developers~\cite{sodhiQuantumComputingPlatforms2021}.
These studies motivate work on testing quantum computing platforms.

Prior to our work, there has been only one other approach on testing quantum computing platforms~\cite{wangQDiffDifferentialTesting2021}, which is a differential testing technique.
In contrast to our work, QDiff does not generate programs from scratch, but starts from six hand-written programs.
Moreover, QDiff performs differential testing across different backends and optimization levels of quantum computing platforms, whereas our work is based on a novel set of metamorphic transformations, only two of which (\textit{change of optimization level} and \textit{change of backend}) are similar to QDiff.
Section~\ref{sec:eval qdiff} empirically shows that \methodName{} reveals bugs missed by QDiff and reaches higher code coverage.

\paragraph{Testing and manipulating quantum programs}
Several approaches for testing quantum programs have been proposed, including a search-based techniques~\cite{wangGeneratingFailingTest2021,wangPosterFuzzTesting2021},
statistical assertion checks that try to limit the effects on the actual computation~\cite{huangStatisticalAssertionsValidating2019, liProjectionbasedRuntimeAssertions2020, liuQuantumCircuitsDynamic2020},
combinatorial testing~\cite{wangApplicationCombinatorialTesting2021},
and coverage-based methods~\cite{aliAssessingEffectivenessInput2021a}.
In contrast to our work, these techniques test specific programs, not the underlying platform.
CutQC~\cite{tangCutQCUsingSmall2021} breaks a quantum circuit into smaller parts so that the resulting sub-circuits can be executed on the limited NISQ devices~\cite{preskillQuantumComputingNISQ2018}.
Our \textit{partitioned execution} transformation also splits a circuit into sub-circuits, but only when the qubits are not entangled, whereas CutQC handles entanglement by approximating the output distribution.

\paragraph{Testing of probabilistic systems}
ProbFuzz~\cite{duttaTestingProbabilisticProgramming2018} is a testing technique targeted at probabilistic systems, such as probabilistic modeling libraries.
While both those libraries and quantum computing platforms output probabilistic distributions, the latter is more deeply connected to hardware constraints, e.g., via a coupling map and the gate set, which our approach considers.

\paragraph{Testing compilers and other developer tools}

The critical role of compilers for overall software reliability has motivated a stream of work on compiler testing.
We refer to a recent survey~\cite{chenEmpiricalComparisonCompiler2016} for a comprehensive overview.
Quantum computing platforms play a similarly critical role in the quantum computing domain, which motivates our work.
Our program generator relates to work on generating traditional programs, e.g., via randomized code generation combined with static and dynamic checks to avoid undefined behavior~\cite{yangFindingUnderstandingBugs2011}, code fragment-based fuzzing~\cite{Holler2012}, and systematic program enumeration~\cite{Zhang2017}.
Metamorphic testing~\cite{seguraSurveyMetamorphicTesting2016} has also been applied in compiler testing, e.g., by deleting and inserting code in the dead regions of a program~\cite{le2014compiler,Le2015}, and via domain-specific transformations for graphics shading compilers~\cite{donaldson2017automated}.
Other developer tools, e.g., debuggers, can also be subject to metamorphic testing~\cite{tolksdorfInteractiveMetamorphicTesting2019}.
None of the above approaches addresses the unique challenges of quantum computing platforms, for which \methodName{} contributes a novel program generator and a novel set of metamorphic transformations.

\section{Conclusion}

Motivated by the increasing popularity of quantum computing, paired with the slim portfolio of techniques for testing its software stack, this paper presents the first metamorphic testing approach for quantum computing platforms.
Our two key contributions are a program generator that efficiently creates a diverse set of non-crashing quantum programs, and a novel set of metamorphic transformations to create pairs of programs to compare with each other.
Our evaluation shows \methodName{}'s effectiveness, e.g., in the form of \nBugReports{} detected bugs in Qiskit.
We envision our contributions to enable future work beyond \methodName{}.
For example, the program generator provides a starting point for other testing techniques, e.g., coverage-guided fuzzing, and the metamorphic transformations could be adapted to other platforms.
Overall, the presented work takes an important step toward further increasing the reliability of software in this still young field.

\section*{Data Availability}
Our implementation and all experimental results are freely and permanently\footnote{Data: \url{https://figshare.com/s/dd0d4af20fd6e06148a3} and software: \url{https://zenodo.org/record/7575881}} available:
\begin{center}
  \url{https://github.com/sola-st/MorphQ-Quantum-Qiskit-Testing-ICSE-23}
\end{center}

\section*{Acknowledgements}
This work was supported by the European Research Council (ERC, grant agreement 851895), and by the German Research Foundation within the ConcSys and DeMoCo projects.

\bibliographystyle{IEEEtran}
\bibliography{reduce_bibliography}

\begin{thebibliography}{10}
\providecommand{\url}[1]{#1}
\csname url@samestyle\endcsname
\providecommand{\newblock}{\relax}
\providecommand{\bibinfo}[2]{#2}
\providecommand{\BIBentrySTDinterwordspacing}{\spaceskip=0pt\relax}
\providecommand{\BIBentryALTinterwordstretchfactor}{4}
\providecommand{\BIBentryALTinterwordspacing}{\spaceskip=\fontdimen2\font plus
\BIBentryALTinterwordstretchfactor\fontdimen3\font minus
  \fontdimen4\font\relax}
\providecommand{\BIBforeignlanguage}[2]{{%
\expandafter\ifx\csname l@#1\endcsname\relax
\typeout{** WARNING: IEEEtran.bst: No hyphenation pattern has been}%
\typeout{** loaded for the language `#1'. Using the pattern for}%
\typeout{** the default language instead.}%
\else
\language=\csname l@#1\endcsname
\fi
#2}}
\providecommand{\BIBdecl}{\relax}
\BIBdecl

\bibitem{paltenghiBugsQuantumComputing2022}
M.~Paltenghi and M.~Pradel, ``Bugs in {{Quantum}} computing platforms: An
  empirical study,'' \emph{Proceedings of the ACM on Programming Languages},
  vol.~6, no. OOPSLA1, pp. 86:1--86:27, Apr. 2022.

\bibitem{Barr2015}
E.~T. Barr, M.~Harman, P.~McMinn, M.~Shahbaz, and S.~Yoo, ``The oracle problem
  in software testing: {A} survey,'' \emph{{IEEE} Trans. Software Eng.},
  vol.~41, no.~5, pp. 507--525, 2015.

\bibitem{Chen1998}
T.~Y. Chen, S.~C. Cheung, and S.~M. Yiu, ``Metamorphic testing: a new approach
  for generating next test cases,'' Technical Report HKUST-CS98-01, Department
  of Computer Science, Hong Kong, Tech. Rep., 1998.

\bibitem{chenMetamorphicTestingReview2018}
T.~Y. Chen, F.-C. Kuo, H.~Liu, P.-L. Poon, D.~Towey, T.~H. Tse, and Z.~Q. Zhou,
  ``Metamorphic {{Testing}}: {{A Review}} of {{Challenges}} and
  {{Opportunities}},'' \emph{ACM Computing Surveys}, Jan. 2018.

\bibitem{QiskitQiskit2021}
``Qiskit/qiskit,'' https://github.com/Qiskit/qiskit, Oct. 2021.

\bibitem{chenSurveyCompilerTesting2020}
J.~Chen, J.~Patra, M.~Pradel, Y.~Xiong, H.~Zhang, D.~Hao, and L.~Zhang, ``A
  {{Survey}} of {{Compiler Testing}},'' \emph{ACM Computing Surveys}, vol.~53,
  no.~1, pp. 1--36, May 2020.

\bibitem{wangQDiffDifferentialTesting2021}
J.~Wang, Q.~Zhang, G.~H. Xu, and M.~Kim, ``{{QDiff}}: {{Differential Testing}}
  of {{Quantum Software Stacks}},'' in \emph{2021 36th {{IEEE}}/{{ACM
  International Conference}} on {{Automated Software Engineering}} ({{ASE}})},
  Nov. 2021, pp. 692--704.

\bibitem{mendiluzeMuskitMutationAnalysis2021}
E.~Mendiluze, S.~Ali, P.~Arcaini, and T.~Yue, ``Muskit: {{A Mutation Analysis
  Tool}} for {{Quantum Software Testing}},'' 2021.

\bibitem{zhaoBugs4QBenchmarkReal2021a}
P.~Zhao, J.~Zhao, Z.~Miao, and S.~Lan, ``{{Bugs4Q}}: {{A Benchmark}} of {{Real
  Bugs}} for {{Quantum Programs}},'' in \emph{2021 36th {{IEEE}}/{{ACM
  International Conference}} on {{Automated Software Engineering}} ({{ASE}})},
  Nov. 2021.

\bibitem{zhaoIdentifyingBugPatterns2021a}
P.~Zhao, J.~Zhao, and L.~Ma, ``Identifying {{Bug Patterns}} in {{Quantum
  Programs}},'' in \emph{2021 {{IEEE}}/{{ACM}} 2nd {{International Workshop}}
  on {{Quantum Software Engineering}} ({{Q-SE}})}, Jun. 2021, pp. 16--21.

\bibitem{fingerhuthOpenSourceSoftware2018}
M.~Fingerhuth, T.~Babej, and P.~Wittek, ``Open source software in quantum
  computing,'' \emph{PLOS ONE}, vol.~13, no.~12, p. e0208561, Dec. 2018.

\bibitem{schuldIntroductionQuantumMachine2015}
M.~Schuld, I.~Sinayskiy, and F.~Petruccione, ``An introduction to quantum
  machine learning,'' \emph{Contemporary Physics}, vol.~56, no.~2, pp.
  172--185, Apr. 2015.

\bibitem{schuldQuestQuantumNeural2014}
------, ``The quest for a {{Quantum Neural Network}},'' \emph{Quantum
  Information Processing}, Nov. 2014.

\bibitem{crossOpenQuantumAssembly2017}
A.~W. Cross, L.~S. Bishop, J.~A. Smolin, and J.~M. Gambetta, ``Open {{Quantum
  Assembly Language}},'' \emph{arXiv:1707.03429 [quant-ph]}, Jul. 2017.

\bibitem{preskillQuantumComputingNISQ2018}
J.~Preskill, ``Quantum {{Computing}} in the {{NISQ}} era and beyond,''
  \emph{Quantum}, vol.~2, p.~79, Aug. 2018.

\bibitem{williamsQuantumGates2011}
C.~P. Williams, ``Quantum {{Gates}},'' in \emph{Explorations in {{Quantum
  Computing}}}, ser. Texts in {{Computer Science}}, C.~P. Williams, Ed.\hskip
  1em plus 0.5em minus 0.4em\relax {London}: {Springer}, 2011, pp. 51--122.

\bibitem{chenEmpiricalComparisonCompiler2016}
J.~Chen, W.~Hu, D.~Hao, Y.~Xiong, H.~Zhang, L.~Zhang, and B.~Xie, ``An
  empirical comparison of compiler testing techniques,'' in \emph{Proceedings
  of the 38th {{International Conference}} on {{Software Engineering}}}, ser.
  {{ICSE}} '16.\hskip 1em plus 0.5em minus 0.4em\relax {New York, NY, USA}:
  {Association for Computing Machinery}, May 2016, pp. 180--190.

\bibitem{kolmogorovSullaDeterminazioneEmpirica1933}
A.~L. KOLMOGOROV, ``Sulla determinazione empirica di una legge di
  distribuzione,'' \emph{G. Ist. Ital. Attuari}, vol.~4, pp. 83--91, 1933.

\bibitem{smirnovTableEstimatingGoodness1948}
N.~Smirnov, ``Table for {{Estimating}} the {{Goodness}} of {{Fit}} of
  {{Empirical Distributions}},'' \emph{The Annals of Mathematical Statistics},
  vol.~19, no.~2, pp. 279--281, Jun. 1948.

\bibitem{zellerIsolatingCauseeffectChains2002}
A.~Zeller, ``Isolating cause-effect chains from computer programs,'' \emph{ACM
  SIGSOFT Software Engineering Notes}, vol.~27, no.~6, pp. 1--10, Nov. 2002.

\bibitem{sivarajahKetRetargetableCompiler2020}
S.~Sivarajah, S.~Dilkes, A.~Cowtan, W.~Simmons, A.~Edgington, and R.~Duncan,
  ``T|ket{$\rangle$}: A retargetable compiler for {{NISQ}} devices,''
  \emph{Quantum Science and Technology}, vol.~6, no.~1, p. 014003, Nov. 2020.

\bibitem{developersCirq2021}
C.~Developers, ``Cirq,'' Zenodo, Aug. 2021.

\bibitem{sodhiQuantumComputingPlatforms2021}
B.~Sodhi and R.~Kapur, ``Quantum {{Computing Platforms}}: {{Assessing}} the
  {{Impact}} on {{Quality Attributes}} and {{SDLC Activities}},'' in \emph{2021
  {{IEEE}} 18th {{International Conference}} on {{Software Architecture}}
  ({{ICSA}})}, Mar. 2021, pp. 80--91.

\bibitem{wangGeneratingFailingTest2021}
X.~Wang, P.~Arcaini, T.~Yue, and S.~Ali, ``Generating {{Failing Test Suites}}
  for~{{Quantum Programs With Search}},'' in \emph{Search-{{Based Software
  Engineering}}}, ser. Lecture {{Notes}} in {{Computer Science}}, U.-M.
  O'Reilly and X.~Devroey, Eds.\hskip 1em plus 0.5em minus 0.4em\relax {Cham}:
  {Springer International Publishing}, 2021, pp. 9--25.

\bibitem{wangPosterFuzzTesting2021}
J.~Wang, F.~Ma, and Y.~Jiang, ``Poster: {{Fuzz Testing}} of {{Quantum
  Program}},'' in \emph{2021 14th {{IEEE Conference}} on {{Software Testing}},
  {{Verification}} and {{Validation}} ({{ICST}})}, Apr. 2021, pp. 466--469.

\bibitem{huangStatisticalAssertionsValidating2019}
Y.~Huang and M.~Martonosi, ``Statistical {{Assertions}} for {{Validating
  Patterns}} and {{Finding Bugs}} in {{Quantum Programs}},'' May 2019.

\bibitem{liProjectionbasedRuntimeAssertions2020}
G.~Li, L.~Zhou, N.~Yu, Y.~Ding, M.~Ying, and Y.~Xie, ``Projection-based runtime
  assertions for testing and debugging {{Quantum}} programs,''
  \emph{Proceedings of the ACM on Programming Languages}, vol.~4, no. OOPSLA,
  pp. 150:1--150:29, Nov. 2020.

\bibitem{liuQuantumCircuitsDynamic2020}
J.~Liu, G.~T. Byrd, and H.~Zhou, ``Quantum {{Circuits}} for {{Dynamic Runtime
  Assertions}} in {{Quantum Computation}},'' in \emph{Proceedings of the
  {{Twenty-Fifth International Conference}} on {{Architectural Support}} for
  {{Programming Languages}} and {{Operating Systems}}}, ser. {{ASPLOS}}
  '20.\hskip 1em plus 0.5em minus 0.4em\relax {New York, NY, USA}: {Association
  for Computing Machinery}, Mar. 2020, pp. 1017--1030.

\bibitem{wangApplicationCombinatorialTesting2021}
X.~Wang, P.~Arcaini, T.~Yue, and S.~Ali, ``Application of {{Combinatorial
  Testing}} to {{Quantum Programs}},'' in \emph{2021 {{IEEE}} 21st
  {{International Conference}} on {{Software Quality}}, {{Reliability}} and
  {{Security}} ({{QRS}})}, Dec. 2021, pp. 179--188.

\bibitem{aliAssessingEffectivenessInput2021a}
S.~Ali, P.~Arcaini, X.~Wang, and T.~Yue, ``Assessing the {{Effectiveness}} of
  {{Input}} and {{Output Coverage Criteria}} for {{Testing Quantum
  Programs}},'' in \emph{2021 14th {{IEEE Conference}} on {{Software Testing}},
  {{Verification}} and {{Validation}} ({{ICST}})}, Apr. 2021, pp. 13--23.

\bibitem{tangCutQCUsingSmall2021}
W.~Tang, T.~Tomesh, M.~Suchara, J.~Larson, and M.~Martonosi, ``{{CutQC}}: Using
  small {{Quantum}} computers for large {{Quantum}} circuit evaluations,'' in
  \emph{Proceedings of the 26th {{ACM International Conference}} on
  {{Architectural Support}} for {{Programming Languages}} and {{Operating
  Systems}}}, ser. {{ASPLOS}} 2021.\hskip 1em plus 0.5em minus 0.4em\relax {New
  York, NY, USA}: {Association for Computing Machinery}, Apr. 2021, pp.
  473--486.

\bibitem{duttaTestingProbabilisticProgramming2018}
S.~Dutta, O.~Legunsen, Z.~Huang, and S.~Misailovic, ``Testing probabilistic
  programming systems,'' in \emph{Proceedings of the 2018 26th {{ACM Joint
  Meeting}} on {{European Software Engineering Conference}} and {{Symposium}}
  on the {{Foundations}} of {{Software Engineering}}}, ser. {{ESEC}}/{{FSE}}
  2018.\hskip 1em plus 0.5em minus 0.4em\relax {New York, NY, USA}:
  {Association for Computing Machinery}, Oct. 2018, pp. 574--586.

\bibitem{yangFindingUnderstandingBugs2011}
X.~Yang, Y.~Chen, E.~Eide, and J.~Regehr, ``Finding and understanding bugs in
  {{C}} compilers,'' \emph{ACM SIGPLAN Notices}, vol.~46, no.~6, pp. 283--294,
  Jun. 2011.

\bibitem{Holler2012}
C.~Holler, K.~Herzig, and A.~Zeller, ``Fuzzing with code fragments.'' in
  \emph{USENIX Security Symposium}, 2012, pp. 445--458.

\bibitem{Zhang2017}
Q.~Zhang, C.~Sun, and Z.~Su, ``Skeletal program enumeration for rigorous
  compiler testing,'' in \emph{PLDI}, 2017.

\bibitem{seguraSurveyMetamorphicTesting2016}
S.~Segura, G.~Fraser, A.~B. Sanchez, and A.~{Ruiz-Cort{\'e}s}, ``A {{Survey}}
  on {{Metamorphic Testing}},'' \emph{IEEE Transactions on Software
  Engineering}, Sep. 2016.

\bibitem{le2014compiler}
V.~Le, M.~Afshari, and Z.~Su, ``Compiler validation via equivalence modulo
  inputs,'' \emph{ACM Sigplan Notices}, vol.~49, no.~6, pp. 216--226, 2014.

\bibitem{Le2015}
V.~Le, C.~Sun, and Z.~Su, ``Finding deep compiler bugs via guided stochastic
  program mutation,'' in \emph{Proceedings of the 2015 ACM SIGPLAN
  International Conference on Object-Oriented Programming, Systems, Languages,
  and Applications}, ser. OOPSLA 2015.\hskip 1em plus 0.5em minus 0.4em\relax
  ACM, 2015, pp. 386--399.

\bibitem{donaldson2017automated}
A.~F. Donaldson, H.~Evrard, A.~Lascu, and P.~Thomson, ``Automated testing of
  graphics shader compilers,'' \emph{Proceedings of the ACM on Programming
  Languages}, vol.~1, no. OOPSLA, pp. 1--29, 2017.

\bibitem{tolksdorfInteractiveMetamorphicTesting2019}
S.~Tolksdorf, D.~Lehmann, and M.~Pradel, ``Interactive metamorphic testing of
  debuggers,'' in \emph{Proceedings of the 28th {{ACM SIGSOFT International
  Symposium}} on {{Software Testing}} and {{Analysis}}}.\hskip 1em plus 0.5em
  minus 0.4em\relax {New York, NY, USA}: {Association for Computing Machinery},
  Jul. 2019, pp. 273--283.

\end{thebibliography}

\end{document}